\documentclass[dvips]{acta}
\usepackage{supertabular,lscape,epsfig}
\usepackage{amssymb}
\usepackage{amsmath}

\SetPages{1}{28}

\SetVol{58}{2008}

\begin{document}

\begin{Titlepage}
\Title{Double-mode classical Cepheid models -- revisited}

\Author{S~m~o~l~e~c, R. and M~o~s~k~a~l~i~k, P.}{Copernicus Astronomical Center, ul.~Bartycka~18, 00-716~Warszawa, Poland\\
e-mail: (smolec,pam)@camk.edu.pl}

\Received{Month Day, Year}
\end{Titlepage}

\Abstract{For many years modeling of double-mode pulsation of classical pulsators was a challenging problem. Inclusion of turbulent convection into pulsation hydrocodes finally led to stable double-mode models. However, it was never analysed, which factor of turbulent convection is crucial. We show that the double-mode behaviour displayed in the computed models results from incorrect assumptions adopted in some of the pulsation hydrocodes, namely from the neglect of buoyant forces in convectively stable layers. This leads to significant turbulent energies and consequently to strong eddy-viscous damping in deep, convectively stable layers of the model. Resulting differential reduction of fundamental and first overtone amplitudes favours the occurrence of double-mode pulsation. Once buoyant forces in convectively stable regions are taken into account (as they should), no stable double-mode behaviour is found. The problem of modeling double-mode behaviour of classical pulsators remains open.}{hydrodynamics -- convection -- overshooting -- stars: oscillations  -- stars: variables: Cepheids -- methods: numerical}

\section{Introduction}%1

Among classical pulsators (Cepheids and RR~Lyrae stars) double-mode pulsators (or beat pulsators) represent a small, but very interesting group. These stars pulsate simultaneously in two, low-order radial modes of pulsation, either in fundamental and first overtone (F/O1) or in two lowest overtones (O1/O2). Their astrophysical importance results from the fact, that two radial modes of pulsation simultaneously observed strongly constrain the physical parameters of the star. More than 30 years ago Petersen (1973) showed, that simple linear theory may be used to calculate the masses of such pulsators. Disagreement between these so-called beat masses and masses derived from evolutionary calculations was one of the factors motivating revision of the opacity tables (Simon 1982). Indeed, with improved physics, new opacity tables solved the beat-mass discrepancy problem (Moskalik, Buchler \& Marom 1992). Beat pulsators proved to be a powerful diagnostic tool for our knowledge of physical processes acting in stars. Recently Kov\'acs (2000a,b) used double-mode Cepheids and RR~Lyrae stars to derive the distance modulus to the Magellanic Clouds. More detailed study of the Petersen diagram for double-mode RR~Lyrae stars in Magellanic Clouds was performed by Popielski, Dziembowski \& Cassisi (2000). Beaulieu \etal (2006) used beat Cepheids to study the metallicity distribution in M33 galaxy. Their methods were further developed by Buchler \& Szab\'o (2007), who show how the metallicity of beat Cepheid may be constrained, using the Petersen diagram.

Results mentioned in the previous paragraph, are based on computations done with linear pulsation hydrocodes. In these codes stability of the model against small perturbations is studied. As a result the pulsation eigenmodes, their periods and growth rates are calculated. Necessary condition for beat pulsation to occur is that the two pulsation modes are simultaneously linearly unstable, that is their linear growth rates, $\gamma$, are positive. For F/O1 beat pulsations one requires that $\gamma_0>0$ and $\gamma_1>0$. These conditions are satisfied in a wide range of astrophysical parameters, covering a significant part of the instability strip. However, the relative scarcity of the beat pulsators in comparison to single-mode pulsators indicate, that the double-mode pulsation domain is much smaller, and is determined by nonlinear effects. Nonlinear computations may further constrain the astrophysical parameters of the beat pulsators. Therefore, from the very beginning of nonlinear pulsation computations, modeling of the beat phenomenon was one of the major objectives. 

Early attempts to model the beat phenomenon with direct time integration, radiative hydrocodes were inconclusive. Limited computer resources didn't allow to perform long-lasting integrations. Mixed-mode states persisting for tens of periods were observed, however, their permanent double-mode nature could not be confirmed. Introduction of relaxation technique (Baker \& von Sengbush 1969, Stellingwerf 1974) greatly facilitated the search for double-mode phenomenon. Relaxation scheme allows for fast convergence to a limit cycle (monoperiodic, finite-amplitude oscillation) and additionally provides information about limit cycle stability through the Floquet stability coefficients. Two stability coefficients are of interest, when one searches for F/O1 double-mode pulsation. $\eta_{0,1}$ measures the stability of the first overtone limit cycle against the fundamental mode perturbation (switching rate toward fundamental mode), while $\eta_{1,0}$ measures the stability of the fundamental mode limit cycle against the first overtone perturbation (switching rate toward first overtone). Positive value of a Floquet coefficient means, that the respective mode is unstable and will switch into the other mode. If both $\eta_{0,1}$ and $\eta_{1,0}$ are positive, double-mode state is unavoidable.

Using relaxation technique several surveys of radiative models were performed to search for the beat pulsation. Examples of resonant, F/O1 doubly-periodic, however, triple-mode models were found by Kov\'acs \& Buchler (1988) (RR~Lyrae models), Buchler, Moskalik \& Kov\'acs (1990) and Smolec (2008) ($\delta$~Cephei models). In all these cases, origin of the doubly-periodic pulsation is connected with the 2:1 resonance between the linearly unstable fundamental mode, and linearly damped higher order overtone (third overtone in case of RR~Lyrae models and second overtone in case of $\delta$~Cephei models). Physical reasons for resonant doubly-periodic pulsation were analysed by Dziembowski \& Kov\'acs (1984) with the help of the amplitude equations formalism. Linearly damped, resonant overtone acts as an energy sink for the linearly unstable fundamental mode, leading to the decrease of its amplitude. As a result, otherwise dominant fundamental mode, is no longer able to saturate the pulsation instability on its own. This allows the growth of the first overtone mode. Due to nonlinear phase-lock, resulting triple-mode pulsation is doubly-periodic. Described resonant F/O1 doubly-periodic models are mainly of theoretical interest, as they do not fulfill the observational constraints. 
  
The search for non-resonant F/O1 (or O1/O2) double-mode pulsation with radiative codes failed, although partial success was achieved by Kov\'acs \& Buchler (1993). They obtained several non-resonant double-mode RR~Lyrae models, by playing with artificial viscosity dissipation. Their models however, are sensitive to numerical details, and do not satisfy all observational constraints. Experiments with artificial viscosity motivated the development of less dissipative codes with physical dissipation instead of artificial one.

Inclusion of turbulent convection into pulsational hydrocodes finally led to robust double-mode behaviour in Cepheid (Koll\'ath \etal 1998) and in RR~Lyrae (Feuchtinger 1998) models. More detailed surveys of the double-mode pulsation models were performed by the Florida-Budapest group (Koll\'ath \& Buchler 2001, Koll\'ath \etal 2002, Szab\'o, Koll\'ath \& Buchler 2004). These analysis showed, that the occurrence of the double-mode behaviour does not depend strongly on the parameters of the convection model used, and is mainly controlled by physical parameters of the stellar model. Detailed comparison with observed double-mode pulsators was not performed, however. Also, particular effect of turbulent convection, responsible for the occurrence of double-mode phenomenon, was not identified. 

In our recent paper (Smolec \& Moskalik 2008, hereafter Paper~I), we described convective hydrocodes we have recently developed. In our hydrocodes, we adopt convection model based on Kuhfu\ss{} (1986) work. Essentially the same convection model was used by Florida-Budapest group, however with modified treatment of the convectively stable layers. In these layers buoyant forces were neglected (Koll\'ath \etal 2002), which is unphysical. Also, convective flux was neglected in convectively stable regions. In Paper~I we showed the consequences of these neglects for the single-mode Cepheid models. In the present paper we show dramatic consequences for the double-mode models.  

The structure of this paper is as follows. In Section~2 we briefly discuss the turbulent convection models we use in this paper. In Section~3 we describe the methods of modal selection analysis we adopt, and discuss typical scenarios for two convection models considered, with buoyant forces present and buoyant forces neglected in convectively stable zones. Reasons for the occurrence of double-mode behaviour in the latter case are explained in Section~4. In Section~5 we describe the extensive survey of models we have computed in search for the double-mode behaviour with convection model including buoyant forces in convectively stable regions. We discuss our results and conclusions in Section~6. 

\section{Turbulent convection models}%2

Modeling of turbulent convection and its coupling with radial pulsation is a very difficult problem. However, its essential features may be captured with simple one-equation models, suitable to implement in nonlinear pulsation hydrocodes. Review of such one-equation models is provided by Baker (1987). Of these, the most consistent one is the model of Kuhfu\ss{} (1986). This model is implemented in our pulsation hydrocodes (Paper I), and very similar model was adopted by the Florida-Budapest group (Yecko \etal 1998, Koll\'ath \etal 2002, hereafter KBSC). In these models generation of turbulent energy, $e_t$, is described by single equation of the following form
\begin{equation}\frac{d e_t}{d t}+(P_t+P_\nu)\frac{d V}{d t}=-\frac{1}{\rho}\frac{\partial(R^2F_t)}{R^2\partial R}+C.\end{equation}
$R$ is radius, $V$ is specific volume, $V=1/\rho$, $F_t$ is turbulent kinetic energy flux, $P_t$ and $P_\nu$ are turbulent and eddy-viscous pressures, respectively, and $C$ is the coupling term of the following form
\begin{equation}C=S-D-D_r.\end{equation}
$S$ is the turbulent source (driving) function, $D$ is turbulent dissipation term and $D_r$ is radiative cooling term. Coupling term couples the turbulent energy equation with thermal energy equation. The latter equation, as well as the equation of motion, are supplemented with additional convection-related terms (convective/turbulent fluxes, pressure terms). Their form is following
\begin{equation}\frac{d U}{d t}=-\frac{1}{\rho}\frac{\partial}{\partial R}(P+P_t+P_\nu)-\frac{GM_R}{R^2},\end{equation}
\begin{equation}\frac{d E}{d t}+P\frac{d V}{d t}=-\frac{1}{\rho}\frac{\partial[R^2(F_r+F_c)] }{R^2\partial R}-C.\end{equation}
$P$ is gas+radiation pressure, $E$ is thermal energy (gas+radiation), $M_R$ is mass enclosed in radius $R$. $F_c$ is convective flux. $F_r$ is radiative flux, which we treat in the diffusion approximation. In the above equations we use eddy viscosity in the form of eddy-viscous pressure, just as was done in the Florida-Budapest code. In original Kuhfu\ss{} model, eddy-viscous terms have different form (see Paper I). However, as we compare our models to those computed by the Florida-Budapest group, we decided to use the same eddy viscosity representation. Results and conclusions presented in this paper do not depend on the exact form of eddy viscosity used in the models (see also Paper I). Convection model contains eight order of unity free parameters (alphas), that should be adjusted to match the observational constraints (in Table~1 we provide the values of $\alpha$ parameters used in the present analysis). We refer the reader to Paper I for the detailed description of the turbulent convection models we use in this paper. 

As we argue in the following, the crucial quantity for understanding the double-mode behaviour is the turbulent source function. Source function describes buoyant forces. Specifically, in the framework of mixing-length theory it may be shown, that source function is proportional to acceleration of convective eddies caused by buoyant forces (Paper~I). In Kuhfu\ss{} theory $S$ is given by
\begin{equation}S=\alpha\alpha_s\frac{TPQ}{H_P}e_t^{1/2}Y,\end{equation}
where $Q$ is thermal expansion coefficient, $H_P$ is pressure scale height and $Y=\nabla-\nabla_\mathrm{a}$ is superadiabatic gradient. $\alpha$ is adjustable, mixing-length parameter, just as $\alpha_s$ parameter, which regulates the strength of the source term. Source function, as well as convective flux, are proportional to superadiabatic gradient, $Y$. According to Schwarzschild-Ledoux criterion, convective instability arises if $Y>0$. The great advantage of the Kuhfu\ss{} model is, that in convectively stable regions, $Y<0$, source term is negative and damps the convective motions (Kuhfu\ss{} 1986, Gehmeyr \& Winkler 1992a,b). Also, in these regions convective flux is negative. Symbolically we write $S\sim Y$ and $F_c\sim Y$. This treatment is adopted in our hydrocode, by default. Different treatment of convectively stable zones was adopted in Florida-Budapest code. Source function and convective flux were restricted to positive values only, and were equal to zero in convectively stable regions. Therefore, we write $S\sim Y_+$ and $F_c\sim Y_+$ for their approach. Our hydrocodes adopt such treatment as an option. Following the convention of Paper I we will refer to the discussed convection models as to NN model ($S\sim Y$ and $F_c\sim Y$) and PP model ($S\sim Y_+$ and $F_c\sim Y_+$). The same notation will be used for Cepheid models computed with these convection recipes. 

In Paper I, we argued that the neglect of buoyant forces in convectively stable layers of PP models is physically not correct. We showed several consequences of this unphysical assumption. Specially, we showed that: ({\it i}) neglect of source function (buoyant forces) in convectively stable regions leads to high turbulent energies below the envelope convective zones \ie in convectively stable layers of the model. These energies are generated at the cost of pulsations through the eddy viscous forces. Therefore, we called this effect artificial overshooting. ({\it ii}) the range of such artificial overshooting is very high, significant turbulent energies extending to more than 6 local pressure scale heights below the envelope convective zones. ({\it iii}) significant turbulent energies in the deep, convectively stable interior of the model lead to strong eddy-viscous damping there. {\it (iv)} consequently, amplitudes of the PP models are smaller than the amplitudes of the NN models, in which buoyant forces are taken into account. ({\it v}) above conclusions are valid independently of the exact values of the convective alpha parameters. Specially, artificial overshooting is present even in models without turbulent flux (in Kuhfu\ss{} theory, turbulent flux accounts for physical overshooting).

In this paper we show further dramatic differences between PP and NN models, concerning the modal selection, specially the double-mode phenomenon. 

\section{Modal selection in PP and NN models}%3

In the following Sections we consider in detail the differences in modal selection scenario between PP and NN models. In Section~3.1 we briefly present methods of modal selection analysis we adopt. In Section~3.2 we compare the modal selection along a sequence of Cepheid models of constant luminosity for both PP and NN models. Conclusions, specially the explanation of the double-mode behaviour observed in PP models will be presented in Section~4.

\subsection{Modal selection analysis}%3.1

In principle, one may establish the modal selection along a sequence of convective models through the use of relaxation technique, similar to the radiative case. However, convective hydrodynamical computations (\eg KBSC) show, that stable double-mode solution may coexist with stable fundamental mode solution for the same model (DM/F hysteresis). In this case, Floquet coefficient of the fundamental mode, $\eta_{1,0}$, is negative, despite the fact that stable double-mode pulsation is possible. Based on Floquet coefficients only, one could easily overlook the double-mode solution. Therefore, to study the modal selection in detail, more sophisticated methods are necessary. We adopt the methods developed by the Florida-Budapest group (see \eg KBSC), combining the direct hydrodynamical integrations with amplitude equations formalism. 

Typical example of hydrodynamical integrations is shown in Fig.~1 in $A_0$-$A_1$ amplitude plane. Integrations of the same static model were started with different initial conditions (crosses in Fig.~1), and time evolution of the surface radius displacement, $\delta R/R_0$, was followed through the analytical signal method, providing the time-dependence of the pulsation amplitudes of the individual modes, $A_k(t)$ (see KBSC for details). $R_0$ is equilibrium surface radius of the model. All trajectories run away from the origin ($A_0=0$, $A_1=0$), as for the presented model both fundamental and first overtone modes are linearly unstable. After initial, fast evolution, trajectories evolve slowly along an arc. It is clearly visible that three left trajectories evolve toward first overtone solution ($A_0=0$, first overtone attractor), while four right trajectories evolve toward fundamental mode solution ($A_1=0$, fundamental mode attractor). Stable double-mode state ($A_0\ne 0$, $A_1\ne 0$) is not possible for the discussed model. 

Presented hydrodynamical results may be described by amplitude equations (AEs) formalism (\eg Buchler \& Goupil 1984). AEs describe the temporal evolution of modal amplitudes, through the set of ordinary differential equations. Considering the non-resonant case and interaction of fundamental and first overtone modes only, AEs truncated at quintic terms are following
\begin{equation}\dot{A_0}=(\gamma_0+q_{00}A_0^2+q_{01}A_1^2+r_{00}A_0^4+r_{01}A_1^4+s_0A_0^2A_1^2)A_0,\end{equation}
\begin{equation}\dot{A_1}=(\gamma_1+q_{10}A_0^2+q_{11}A_1^2+r_{10}A_0^4+r_{11}A_1^4+s_1A_0^2A_1^2)A_1.\end{equation}
$\dot{A_0}$ and $\dot{A_1}$ are time derivatives of fundamental and first overtone amplitudes. $\gamma_0$ and $\gamma_1$ are linear growth rates\footnote{In this paper we use radius growth rates, $\gamma$, which differ from kinetic energy growth rates per pulsation cycle, $\eta$, (used in Paper~I) by factor $2P$, where $P$ is period of the mode considered; $2P\gamma=\eta$}, $q_{ij}$ and $q_{ii}$ are cubic, cross- and self-saturation coefficients, $s_0$, $s_1$ and $r_{ij}$ are quintic saturation coefficients. All quantities in the above equations are real, as non-resonant AEs (complex in general) may be decoupled into real part for amplitudes, eqs. (6)-(7), and imaginary part for phases (not important in our considerations, see \eg KBSC). Linear growth rates and saturation coefficients are functions of the model mass, luminosity, effective temperature, chemical composition and physics (\eg opacities, convection). 

In the study of radiative models, AEs were truncated at cubic terms, as these were sufficient to describe the radiative hydrodynamical models. Such cubic AEs are insufficient to capture the richer topology of convective models. In particular, coexistence of stable single-mode and double-mode solutions cannot be reproduced (Buchler \& Kov\'acs 1986). In order to capture DM/F hysteresis quintic terms must be included. However, it is not necessary to retain all quintic terms. Buchler \etal (1999) retain $r$-terms only, while KBSC retain $s$-terms only and claim that  these terms slightly better describe hydrodynamical models. Therefore, we also keep the $s$-terms, and neglect the $r$-terms in our main analysis. 

Time-independent, constant amplitude solutions (fixed points) of the quintic AEs and their stability may be easily calculated if saturation coefficients and linear growth rates are known. For single-mode solutions we have
\begin{equation}A_0=\sqrt{-\gamma_0/q_{00}}, \ \ \ A_1=0,\end{equation}
\begin{equation}A_1=\sqrt{-\gamma_1/q_{11}}, \ \ \ A_0=0,\end{equation}
for the fundamental mode and for the first overtone fixed points, respectively. Linear stability of these solutions with respect to the perturbation in the other mode is described by linear stability coefficients
\begin{equation}\gamma_{1,0}=\gamma_1+q_{10}A_0^2=\gamma_1-\gamma_0\frac{q_{10}}{q_{00}},\end{equation}
\begin{equation}\gamma_{0,1}=\gamma_0+q_{01}A_1^2=\gamma_0-\gamma_1\frac{q_{01}}{q_{11}}.\end{equation}
$\gamma_{1,0}$ describes the stability of the fundamental mode fixed point with respect to the first overtone perturbation, while $\gamma_{0,1}$ describes the stability of the first overtone fixed point. Positive value of the stability coefficient means, that respective, single-mode fixed point is unstable. The system of quintic AEs can also have up to two double-mode fixed points. Analytical expressions for their amplitudes can be given and their stability may be checked through the Hurwitz criteria. Stable fixed points are attractors of the system and trajectories evolve toward them, while unstable fixed points repel the trajectories. Fixed points correspond to steady nonlinear pulsations. Stable single-mode fixed points correspond to limit cycle pulsation, and their linear stability coefficients, $\gamma_{1,0}$ and $\gamma_{0,1}$, are directly related to Floquet coefficients of the limit cycles.

In principle, saturation coefficients may be calculated for a given stellar model, using its static structure and linear eigenvectors. In practice, it is too complicated and may be done only if some simplified assumptions are made. However, hydrodynamical results, such as presented in Fig.~1, allow for straightforward determination of all coefficients entering the AEs, (6)-(7). Analytic signal method (KBSC) provides amplitudes, $A_0(t)$ and $A_1(t)$, and smooth time derivatives, $\dot{A_0}(t)$ and $\dot{A_1}(t)$, at each moment of the model evolution. These data may be used to determine the saturation coefficients and linear growth rates, through simple linear fit. Once coefficients of the AEs are known, all fixed points and their stability may be easily calculated. In Fig.~1 stable fixed points are marked with filled squares (fundamental and first overtone single-mode attractors), while unstable fixed points with open squares (origin and unstable double-mode fixed point). Short line segments in Fig.~1 visualize the flow field, ($\dot{A_0},\dot{A_1}$), computed using the AEs. Flow field is normalized, so the lines show the expected direction of evolution, but not its speed. It is clearly visible in Fig.~1, that results of hydrodynamic computations are well reproduced by the AEs. 

To examine a modal selection along a sequence of Cepheid models with constant luminosity, we repeat the described procedure for series of models differing in effective temperature. For each model three or four trajectories, probing different regions of the $A_0$-$A_1$ plane are calculated and used to derive the saturation coefficients and linear growth rates, through a procedure described above. Then, using interpolation, linear growth rates and saturation coefficients may be computed at any temperature along the sequence, together with all fixed points and their stability. 

Described method allows for detailed study of the modal selection along a sequence of models. It allows to find a possibly very narrow temperature ranges with stable double-mode solution, as pulsation state is calculated for any temperature. It also allows to find a stable double-mode solution coexisting with stable single-mode solution, which is not possible with relaxation technique. On the other hand, the method is time-consuming, as many hydrodynamic computations over many pulsation cycles are necessary. In the next Section we apply the method to NN and PP model sequences.

\subsection{PP models versus NN models}%3.2

In this Section, we compare the modal selection for two sequences of Cepheid models of the same physical parameters ($M$, $L$, $X$, $Z$), convective parameters (alphas) and numerical parameters (mesh construction). The only difference between them is different treatment of the source function and convective flux in convectively stable regions (either PP or NN model). The goal is to verify and explain the double-mode behaviour computed by Florida-Budapest group (PP models), and check, whether similar modal selection is observed in our approach (NN models).

All models considered in this analysis (both PP and NN) are constructed and computed with our hydrocodes, just as described in Paper I. At this point we focus our attention on set A of convective parameters (Table~1). This set represents the simplest convective model without turbulent pressure and turbulent flux ($\alpha_p=\alpha_t=0$).  Models have constant mass, $M=4.5\mathrm{M_\odot}$, constant luminosity, $L=1143.5\mathrm{L_\odot}$, and Galactic chemical composition ($X=0.7$, $Z=0.02$). We use OPAL opacity tables (Iglesias \& Rogers 1996 + Alexander \& Ferguson 1994 at low temperatures) and  solar mixture of Grevesse \& Noels (1993). According to Koll\'ath \& Buchler (2001) for Cepheid sequence of $4.5\mathrm{M_\odot}$, double-mode behaviour should be present in relatively large range of effective temperatures. To avoid confusion, we will precede the set name by NN or PP prefix, depending on the convection model used. Once again we stress, that the PP-A and NN-A models differ only in a treatment of the convectively stable zones.

 Modal selection analysis proceeds just as described in the previous Section. We consider sequence of models with 25K step in temperature, for which both fundamental and first overtone modes are simultaneously unstable. In Fig.~2 we plot the cubic saturation coefficients derived through the linear fit of the amplitude equations to hydrodynamic computations. In Fig.~2a we present the results for set NN-A, while in Fig.~2b for set PP-A. It is clearly visible that saturation coefficients vary smoothly along a temperature sequence. Amplitude equations describe the nonlinear hydrodynamic computations very well in case of both convection models (this is clearly visible in Fig.~1 for NN model). However, linear growth rates are not reproduced equally well for PP and NN convection model, which we describe below.

Linear growth rates may be computed using linear hydrocode (computed growth rates) and, independently, they may be derived through the described fitting procedure (fitted growth rates). Comparing such derived linear growth rates we notice differences, depending on the convection model used and the form of the AEs considered. In case of NN convection model and quintic AEs with $s$-terms only, linear growth rates for the first overtone agree very well, within $\pm 5$ per cent. For growth rates of the fundamental mode, we note a systematic difference. Fitted growth rates are on average higher by $\sim 15$ per cent than computed growth rates. Such systematic difference is not unexpected, since hydrodynamic results are described by truncated series of the amplitude equations. However, we expect, that the difference will be smaller if more terms are retained in AEs. Indeed, when we use seventh order amplitude equations, linear growth rates of the fundamental mode agree to within $\pm 3$ per cent. Differences are much larger for PP convection model. Even if we use seventh order AEs, fitted growth rates are systematically smaller than computed growth rates by roughly 15 per cent for both fundamental and first overtone modes. Most likely, such difference is connected with the not differentiable nature of the PP convection model, where source function and convective flux are truncated in convectively stable zones.  We also note, that in the fitting procedure, we may use computed growth rates, and fit only the saturation coefficients, instead of fitting all coefficients of the AEs. In case of NN models, both approaches (linear growth rates fitted or fixed) provide equally good description of the hydrodynamic computations. In case of PP models, when linear growth rates are not fitted description is worse (specially for models lying close to the edge of the instability strip). We note, that the Florida-Budapest group did not use the computed linear growth rates, but fitted them instead (KBSC, Szab\'o \etal 2004). In the following we also fit the linear growth rates, as such procedure allows for better description of the hydrodynamic results computed with PP convection model.

Knowing linear growth rates and saturation coefficients for some particular temperatures along a model sequence, their values at any temperature may be calculated through simple linear interpolation. Then, for any temperature one may calculate all fixed points and their stability. These results are shown in Figs.~3 for the NN-A model sequence (Fig.~3a) and PP-A model sequence (Fig.~3b). Amplitudes of the single-mode and double-mode fixed points are plotted versus the effective temperature of the models. Solid lines correspond to stable, while dashed lines to unstable solutions. Amplitude of the fundamental mode in the single-mode (SM) as well as in the double-mode (DM) fixed points is plotted with thick line. Amplitude of the first overtone fixed points is plotted with thin line.

Considering the NN-A sequence (Fig.~3a) no stable double-mode solution is possible. At higher temperatures we are close to the fundamental mode blue edge. First overtone is then the only attractor of the system. At temperature $\sim 6290$K fundamental mode limit cycle becomes stable, and unstable double-mode solution appears. In the temperature range $6290$K -- $6165$K there are two stable single-mode attractors and one unstable double-mode solution. At temperature $\sim 6165$K first overtone limit cycle loses its stability and unstable double-mode solution disappears. For lower temperatures, fundamental mode is the only attractor of the system. In the described sequence, first overtone and fundamental mode pulsation domains are separated by the so-called either-or domain, where single-mode pulsation in either modes is possible. We note, that qualitatively the same modal selection scenario is observed in radiative models (\eg Stellingwerf 1975, Buchler \& Kov\'acs 1986).

Considering the PP-A sequence (Fig.~3b) double-mode behaviour is easily found in hydrodynamical models. The double-mode pulsation domain extends to slightly more than $50$K, in the range of $6150$K - $6100$K. For higher temperatures, only first overtone limit cycle is stable, while for lower temperatures, only fundamental mode limit cycle is stable. In this case, no hysteresis state is possible. At each temperature, only one stable pulsation state is possible. 

Described modal selection scenarios for NN and PP models are typical. Qualitatively the same results are obtained for models of set B (Table~1) where effects of turbulent pressure and turbulent flux (convective overshooting) are turned on. In Section~5 we will show, that no stable double-mode solution can be found in NN models, despite extensive search for it. For PP models double-mode solution is easily found. For both PP-A and PP-B sets, double-mode pulsation domain extends to more than 50K. Our more extensive computations with other sets of convective parameters (not presented in this analysis) confirm, that in PP models double-mode behaviour is quite common and its existence does not depend strongly on the values of convective parameters, but is restricted by physical parameters of the models. This was already noticed by KBSC. We do not intend to make any systematic survey of double-mode models with PP convection model, as this, to some extent was done by Florida-Budapest group (Koll\'ath \& Buchler 2001, KBSC). We focus our attention on comparison with NN models, which leads to the understanding of double-mode behaviour in PP models.

\section{Explanation of double-mode behaviour in PP models}%4

Since the first computations of the F/O1 double-mode models with convective hydrocodes (Koll\'ath \etal 1998, Feuchtinger 1998), only Florida-Budapest group performed more extensive surveys of the double-mode pulsations (Koll\'ath \& Buchler 2001, KBSC, Szab\'o \etal 2004). However, the reasons for stable double-mode behaviour were never clearly identified. Using cubic amplitude equations it is straightforward to show (Dziembowski \& Kov\'acs 1984, Buchler \& Kov\'acs 1986), that necessary condition for the double-mode behaviour is positive value of the following determinant, $\mathcal{D}=q_{00}q_{11}-q_{01}q_{10}$, which means that self-saturation is more efficient than cross-saturation. Closer inspection of Fig.~2b reveals that indeed, $\mathcal{D}>0$ for sequence of PP models, although one should remember that discussed condition is not strictly valid, when quintic amplitude equations are considered. KBSC list two reasons for the occurrence of double-mode behaviour in turbulent convection models. First, increase of the self-saturation terms, $q_{00}$, $q_{11}$, relative to cross-saturation terms, $q_{01}$, $q_{10}$. As saturation coefficients are complicated functions of model structure, uncovering the cause of double-mode pulsation is not easy. Second, the necessity to include the quintic terms in the description of hydrodynamical results, which allow to find a hysteresis solution, easy to overlook, if these terms are neglected. This is however, rather claim of the fact, that with turbulent convection more complicated modal selection scenario is possible.

Positive value of $\mathcal{D}$, does not give much insight into physical mechanisms responsible for the double-mode behaviour. Particularly, it was never analysed, which factor of turbulent convection model is necessary for the occurrence of double-mode behaviour. The fact that the occurrence of double-mode behaviour is not sensitive to the exact values of the convective alpha parameters indicate, that double-mode solution is characteristic for PP convection model, as such. 

The reasons for stable double-mode behaviour observed in PP model sequences become clear, when these models are compared with NN models, for which double-mode behaviour is not observed. We focus our attention on just discussed set A, however, it will become clear that the following discussion is general and holds for any set of convective parameters. Corresponding models of sets PP-A and NN-A have the same physical, convective and numerical parameters. Therefore it is obvious, that the dramatic difference in modal selection (Fig.~3a \vs Fig.~3b) is caused by different treatment of convectively stable zones in PP and NN models. As this treatment in PP models is incorrect and leads to unphysical effects (Section~2 of this paper; Paper~I), also double-mode models computed with PP convection model are unphysical.

In Paper I we have shown, that amplitudes of the PP models are lower, than amplitudes of the NN models of the same parameters. This effect is clearly visible in Fig.~4, where we plot the results of hydrodynamical integrations for the same model of set PP-A (trajectories plotted with solid lines) and of set NN-A (trajectories plotted with dotted lines). Difference is striking, amplitudes for PP model being  much smaller. In Paper~I we have explained this effect in detail. Neglect of negative buoyancy effects in PP models leads to relatively high turbulent energies in convectively stable, internal parts of the model (artificial overshooting). Consequently, significant eddy-viscous dissipation in these parts of the model leads to smaller amplitudes in comparison to NN model, in which artificial overshooting is not present. Fig.~4 reveals additional, crucial feature. Amplitude of the fundamental mode is reduced more than the amplitude of the first overtone. Concerning single-mode fixed points, amplitude of the fundamental mode is reduced by factor $\sim3.3$, while amplitude of the first overtone only by factor $\sim1.5$. We conclude, that eddy-viscous damping present in convectively stable zones of PP models, acts differentially on the pulsation modes, having stronger effect on the fundamental mode. It is a direct consequence of different properties of fundamental and first overtone modes in the interior of the model. In the region of significant turbulent energies, below envelope convection zone, first overtone has much smaller pulsational amplitude than the fundamental mode (specially below its pulsation node). Therefore eddy-viscous damping is weaker for the first overtone. This effect is clearly visible in nonlinear eddy-viscous work integrals for both modes, presented in Figs.~5. As for the PP-sequence with stable double-mode solution two single-mode solutions cannot be simultaneously stable, work integrals are plotted for two models of different effective temperature. Fundamental mode eddy-viscous work integral is plotted for model of set PP-A, lying directly to the red of the double-mode pulsation domain ($6075$K) and first overtone work integral is plotted for the model lying directly to the blue of the double-mode pulsation domain ($6175$K). We do not expect significant changes in the work integrals of the respective modes, in a narrow range of temperatures considered here, therefore such comparison is justified. Arrows in Figs.~5 around zone 46, mark the approximate location of the first overtone pulsation node. Arrows around zone 70, mark the bottom boundary of the envelope convective zone. It falls approximately at the same location for both models compared in the Figures. Zones below zone 70 are convectively stable. In these zones, crucial difference in the work integrals is observed. Due to artificial overshooting present in PP models, turbulent energies are significant in the discussed region. Consequently, strong eddy-viscous damping is present for both modes. However, for the first overtone it is insignificant below zone $\approx$50, as pulsation amplitude of the first overtone becomes very small. In the same zones, amplitude of the fundamental mode is still significant and so is eddy-viscous damping. It becomes negligible in the deeper regions, below zone $\approx$30. Above bottom boundary of the envelope convective zone, eddy-viscous damping is similar for both fundamental and first overtone modes. For NN models eddy-viscous damping is not present below the envelope convective zones (Paper~I). Therefore, it is the internal eddy-viscous damping, present in convectively stable zones of PP models, which is responsible for the differential reduction of their pulsation amplitudes, as compared to NN models. 

Differential reduction of modal amplitudes in PP models in comparison to NN models is crucial in bringing up the double-mode behaviour. As is visible in eqs. (10) and (11), amplitude of the single-mode fixed point is the main factor related to its stability. Lower the amplitude of the single-mode solution, more unstable it is against perturbations in the other modes. Comparing PP and NN models (Figs.~3, Fig.~4) we clearly see, that in PP models the amplitude of the fundamental mode is reduced significantly more than the amplitude of the first overtone. Therefore, we expect, that fundamental mode limit cycle is more unstable in PP models. This is clearly visible in Fig.~6a, where we  present the run of stability coefficients, $\gamma_{1,0}$ and $\gamma_{0,1}$, for PP-A sequence (solid lines) and NN-A sequence (dashed lines). Stability coefficient of the fundamental mode fixed point, $\gamma_{1,0}$, is plotted with thick line, while stability coefficient of the first overtone fixed point, $\gamma_{0,1}$, with thin line. This Figure should be analysed together with Figs.~3, showing the amplitudes of the single-mode solutions. Moving across the instability strip to the red, first overtone limit cycle becomes more unstable ($\gamma_{0,1}$ increases), while fundamental mode limit cycle becomes more stable ($\gamma_{1,0}$ decreases). For both PP and NN sequences the run of $\gamma_{0,1}$ is very similar and first overtone limit cycle becomes unstable around temperature of $6150$K. However, dramatic differences are visible in the stability of the fundamental mode limit cycle. For NN sequence, fundamental mode limit cycle becomes stable at relatively high temperature, just as its amplitude becomes higher than the amplitude of the first overtone (Fig.~3a). For lower temperatures its amplitude is much higher, and so it is very stable. Stable double-mode solution is not possible for NN sequence. For PP sequence and higher temperatures, amplitude of the fundamental mode is lower, than the amplitude of the first overtone (Fig.~3b). Fundamental mode limit cycle is unstable at higher temperatures, while first overtone limit cycle is stable. As temperature of the models decrease, amplitude of the fundamental mode grows. First overtone limit cycle becomes unstable and stable double-mode solution appears. Double-mode domain extends from $\sim 6150$K to $\sim 6100$K, where fundamental mode limit cycle becomes stable. This temperature is lower by $\approx 200$K in comparison to NN model sequence.

The mechanism bringing up the double-mode behaviour in PP models, is somewhat analoguous to the mechanism causing the 2:1 resonant triple-mode behaviour. In resonant case, amplitude of one of the two linearly unstable modes is reduced, through its interaction with the linearly damped mode, which acts as an energy sink. As a result, otherwise dominant mode is no longer able to saturate the other linearly unstable mode, which allows its growth. In PP sequence, the internal eddy-viscous damping plays the same role as the resonance - it reduces the amplitude of both modes, however, differentially. The amplitude of the otherwise dominant fundamental mode is reduced much more than the amplitude of the first overtone. It allows the growth of the first overtone and occurrence of stable double-mode solution. Internal eddy-viscous damping, acts as an energy sink for both modes, however, more efficiently for the fundamental mode, as this mode has significantly higher amplitude below envelope convection zone, in the region of significant turbulent energies.

As we discussed, the occurrence of double-mode behaviour is caused by the differential reduction of amplitudes, through the internal eddy-viscous damping. This damping, occurring in convectively stable regions, is unphysical, as we argue in detail in Paper~I, and summarize in Section~2. It is present in all models computed with PP convection model, not depending on the values of convective parameters (Paper~I). In Fig.~6b we present the run of stability coefficients for set B for both PP and NN treatments. Indeed, qualitatively the same picture is observed as for set A (Fig.~6a). Again, with PP convection model both modes are simultaneously unstable in temperature range of $5880$K -- $5820$K, leading to stable double-mode behaviour. For NN-B sequence, fundamental mode limit cycle becomes stable at high temperatures, and is firmly stable in almost whole temperature domain. This is caused by large amplitude of the fundamental mode, which exceeds the amplitude of the first overtone soon after the fundamental mode becomes linearly unstable.

We have also computed additional sequences of models with different sets of convective parameters using PP convection model. We conclude, that no particular ingredient of the PP convection model, such as \eg tur\-bu\-lent pressure, turbulent flux or radiative loses, is crucial in bringing up the double-mode behaviour. It is the neglect of buoyant forces in convectively stable regions, which is responsible for double-mode pulsations.

\section{Search for double-mode phenomenon in NN models}%5

Unphysical mechanism, responsible for the occurrence of double-mode behaviour in PP models, is not present in NN models, where convectively stable regions are treated more properly. However, it does not mean that the double-mode behaviour should be {\it a priori} excluded in NN model sequences, at all. F/O1 double-mode Cepheids do occur in nature, and hydrocode, with correct description of physical processes, such as convection or radiation transfer (opacities), should be able to produce the double-mode behaviour. Therefore, it is very important to search for the double-mode pulsations with NN convection model. This is however, a laborious task. Since double-mode behaviour may be present for some special combination of convective alpha parameters, many sets should be carefully checked. Also, the range of physical parameters, for which double-mode behaviour is possible, is expected to be narrow. As modal selection analysis methods we use, are very time consuming, special strategy must be used for search of the double-mode behaviour. As all models computed in this Section use the NN convection model, we drop the NN prefix before the set name in the following. 

We consider two basic sets of convective parameters, A and B (Table~1)\footnote{Sets A and B considered in this analysis, are identical to sets A and B in Paper~I, except that different form of eddy viscosity is used.}. Set A represents the simplest possible convection model, without turbulent pressure and turbulent flux. These effects are turned on in set B. In all these sets turbulent flux limiter is turned off, and we use eddy viscosity in the form of eddy-viscous pressure (\cf Paper I). Physical parameters for these two sets are the same, namely $M=4.5\MS$, $L=1143.5\LS$, $X=0.07$ and $Z=0.02$. This are typical parameters for the Galactic Cepheids, leading to model periods and period ratios, typical for the observed double-mode Cepheids. In all computations we use OPAL opacity tables and Grevesse \& Noels (1993) solar mixture (except in set A4, where Asplund, Grevesse \& Sauval 2005 mixture is used).

Our search for double-mode behaviour proceed in two steps. In the first step, we keep the physical parameters fixed and vary the convective parameters (Section~5.1), while in the second step, we consider our basic sets of convective parameters, A and B, and vary the physical parameters of the models (Section~5.2). Our model sequences have constant mass and luminosity and varying effective temperature. Models are computed in a narrow range of temperatures (width $\sim 300$-$400$K), where both fundamental and first overtone are linearly unstable, and where transition in modal selection is expected. Using methods described in Section~3.1 we established modal selection for each sequence of models we computed. In no of the computed sequences we find a stable double-mode pulsation. We also tried to select the promising combinations of parameters, for which, double-mode behaviour is most likely. This was done through comparison of computed runs of stability coefficients, $\gamma_{0,1}$ and $\gamma_{1,0}$, for a set considered, with corresponding stability coefficients of our basic sets, A and B (dashed lines in Figs.~6a,b). As described in the previous Section, for both A and B sets of convective parameters, fundamental mode becomes nonlinearly stable soon after becoming linearly unstable, due to its large amplitude. Therefore, the sets for which fundamental mode limit cycle is less stable for higher temperatures are more promising in the context of search for double-mode behaviour. Our unsuccessful efforts to find a double-mode solution are described in the following sections.

\subsection{Effects of varying convective parameters}%5.1

In the upper part of Table~1, we list all sets of convective parameters we have investigated. In sets A1-A8, turbulent flux and turbulent pressure are neglected, and so the results for these sets will be compared to the results for set A. In sets B1 and B2, turbulent pressure and turbulent flux are turned on, and hence the results for these sets will be compared to the results for set B. In any of the discussed sets we haven't found the double-mode behaviour. Tendencies among stability coefficients of the limit cycles are discussed below.

{\it Effects of varying $\alpha$}. In sets A1 and A2 we study the effects of varying mixing-length parameter, $\alpha$. This parameter is present in most of the convective quantities and multiplies (or divides) other alpha-parameters of the model (see Paper~I). By varying its value we increase/decrease the overall strength of convection in the model considered. In basic set A, $\alpha$ is set to $\alpha=1.5$. In set A2 it is increased to $\alpha=1.7$, while in set A1 it is decreased to $\alpha=1.3$. 

With increased $\alpha$ (set A2), we observe that first overtone limit cycle is more unstable in comparison to set A, while fundamental mode limit cycle is more stable. Reverse is true if $\alpha$ is decreased (set A1). As a result the overall situation does not change much in comparison to set A. In both cases, fundamental mode limit cycle becomes firmly stable at high temperatures, preventing the occurrence of the double-mode solution.

{\it Effects of varying $\alpha_m$}. The $\alpha_m$ parameter controls the strength of the eddy-viscous damping present in the model. It may be used to adjust the pulsation amplitudes of the models. Koll\'ath \& Buchler (2001) found, that in case of the PP convection model, modal selection depends on the value of $\alpha_m$. With higher $\alpha_m$ double-mode solution was the only attractor of the system, for some particular range of temperatures, while with lower $\alpha_m$, DM/F hysteresis was observed instead.

In set A3 we slightly increase the strength of the eddy-viscosity by setting $\alpha_m=0.25$. We observe, that first overtone limit cycle is more unstable in comparison to set A. Stability of the fundamental mode limit cycle is similar as observed in set A at lower temperatures, while at higher temperatures, fundamental mode limit cycle is more unstable. This increases the likelihood of the double-mode occurrence. However, when we increase the value of $\alpha_m$, instability strip becomes narrower. Therefore, further increase of $\alpha_m$ would stabilize the first overtone, before double-mode solution can be obtained.

{\it Effects of varying $\alpha_s$, $\alpha_c$ and $\alpha_d$}. In our basic sets, A and B, these parameters are set to their standard values, chosen to reduce the static convection model to standard mixing-length theory (see Paper~I). $\alpha_s$ defines the strength of the source term, $\alpha_c$ regulates the strength of the convective flux, and $\alpha_d$ regulates the strength of the turbulent dissipation. 

In sets A4 and A5 we have decreased the values of $\alpha_s$ and $\alpha_c$ by 20 per cent, in comparison to their standard values, while in set A6 we have increased the value of $\alpha_d$ by 20 per cent in comparison to its standard value. In all sets, A4-A6, we observe qualitatively the same situation: fundamental mode limit cycle becomes more unstable in comparison to set A, while first overtone limit cycle is slightly more stable for lower temperatures, and its stability is similar as in set A for higher temperatures. This behaviour is promising in the context of search for the double-mode pulsations. Therefore, we have computed additional sequence of models with set A7 of convective parameters. Set A7 corresponds to the 'merged' sets A4-A6, as we have $\alpha_s=0.8$, $\alpha_c=0.8$ and $\alpha_d=1.2$ for this set. Again, we observe that fundamental mode limit cycle is more unstable in comparison to set A, however, first overtone limit cycle becomes much more stable, and we are receding from the double-mode solution. 

{\it Effects of radiative losses}. In set A8 we study the effects of radiative cooling of convective elements. This process is described by radiative cooling term, which is additional quantity entering the coupling term (eq.~(2)). We set $\alpha_r$ to its standard value (\cf Paper~I). Models with radiative cooling included in the convection model have significantly higher pulsation amplitudes. Therefore, in set A8 we increase the value of $\alpha_m$ to $\alpha_m=0.3$. Although fundamental mode limit cycle is more unstable in comparison to set A, first overtone limit cycle is firmly stable and double-mode solution is not possible.

{\it Effects of turbulent pressure and turbulent flux}. In sets B, B1 and B2 turbulent pressure and turbulent flux are turned on. In all cases $\alpha_p$ is set to its standard value, and the strength of turbulent flux is varied, through the value of $\alpha_t$ parameter. $\alpha_t$ equals to 0.01, 0.1 and 0.5 in sets B, B1 and B2, respectively. In these sets eddy viscosity parameter is set to $\alpha_m=0.25$ (set B) and to slightly lower value, $\alpha_m=0.20$, in sets B1 and B2.

It is hard to compare the stability of limit cycles in case of models with turbulent pressure and turbulent flux and models without these effects. This is due to the strong shift of the instability strip toward higher temperatures in case of the latter models. However, comparing Figs.~6a~and~6b, we observe qualitatively the same modal selection for sets A and B.

Comparing sets B, B1 and B2, we observe that as strength of turbulent flux is increased by setting the value of $\alpha_t$ to $\alpha_t=0.1$ in set B1 (ten times larger than in set B), stability of both of the modes is almost not affected. Further increase of the $\alpha_t$ value to $\alpha_t=0.5$ (set B2), leads to more stable fundamental mode limit cycle and more unstable first overtone limit cycle, and hence possible double-mode solution becomes more distant.
 
\subsection{Effects of varying physical parameters}%5.2

In the lower part of Table~1, we list all sets of physical parameters we investigated. We studied the effects of varying mass, mass-luminosity relation, and metallicity. Convective parameters were fixed, either to the values of set A (sets  AM, AL1, AL2, AC1-AC4), or to the values of set B (sets BM, BL1, BL2, BC1, BC2). In neither of the model sequences, we have found double-mode behaviour. Below we summarize our results.

{\it Effects of varying mass}. Effects of varying mass (mass-luminosity relation fixed) are studied in sets AM and BM. Mass of the models is set to 4.0\MS, and luminosity is accordingly lower. For lower mass instability strip shifts toward higher temperatures. Considering the stability of the modes in both sets of models, AM and BM, we observe qualitatively the same picture as for sets A and B, respectively. In a large range of temperatures both modes are firmly stable, fundamental mode limit cycle being stable soon after crossing the blue edge of the fundamental mode instability strip. No tendency toward double-mode solution is observed.

{\it Effects of varying luminosity}. In sets AL1, AL2 and BL1, BL2, we study the effects of varying luminosity. We keep the mass fixed to 4.5\MS but decrease the luminosity. In sets AL1 and BL1 it is decreased to 902.5\LS (in comparison to $L=1143.5\LS$ in sets A and B), while in sets AL2 and BL2 we use luminosity resulting from the mass-luminosity relation for the first crossing ($L=404.4\LS$, Alibert \etal 1999).

In both sets with lower luminosity, AL1 and BL1, we observe that the first overtone limit cycle is slightly more unstable in comparison to sets A and B, respectively, specially for higher temperatures. Fundamental mode limit cycle is more unstable for lower temperatures and more stable for higher temperatures. Again, no tendency toward double-mode solution is observed. For sets obeying the first crossing M-L relation (AL2 and BL2) situation is even worse. First overtone limit cycle is firmly stable in a wide range of temperatures. It becomes unstable, far from the blue edge, when fundamental mode limit cycle is also firmly stable.

{\it Effects of chemical composition}. In sets AC1-AC3 and BC1, BC2 we keep the mass and luminosity as in our basic sets, but vary the chemical composition. We set $X=0.716$, $Z=0.1$ ($\sim$ LMC) in sets AC1 and BC1, and $X=0.756$, $Z=0.004$ ($\sim$ SMC) in sets AC2 and BC2. Also in set AC3 we set higher helium abundance ($X=0.726$, $Z=0.004$).

In all sets we observe qualitatively the same behaviour of stability coefficients. First overtone limit cycle becomes unstable at roughly the same temperature in sets A, AC1-AC3 and in sets B, BC1 and BC2 (spread of order of $50$K). At this temperature fundamental mode limit cycle is firmly stable. Lower the metallicity, more stable fundamental mode limit cycle.

We also checked the effects of recent revision of solar chemical composition and solar metallicity (Asplund, Grevesse \& Sauval 2005), by computing additional sequence of models with set A4 of convective parameters. Metallicity is set to $Z=0.012$ and new solar mixture is used. Qualitatively the same modal selection is observed as for set A.

\section{Discussion and Conclusions}%6

We have clearly identified the reasons for stable double-mode behaviour, observed in models ignoring negative buoyancy effects (PP models). Double-mode pulsation results from the neglect of the buoyant forces in convectively stable zones. As a consequence, turbulent velocities are not braked effectively below the envelope convective zone, which leads to artificial overshooting. The range of such artificial overshooting is very large, significant turbulent energies extend to more than 6 local pressure scale heights below the envelope convective zones (Paper~I). These turbulent energies are high enough to produce a significant eddy-viscous damping in the internal convectively stable regions of the model. This damping acts differentially on the pulsation modes, having stronger effect on the fundamental mode than on the first overtone, since the latter has significantly lower amplitude in the deep envelope. Without this damping (in NN models which take negative buoyancy into account) amplitude of the fundamental mode pulsation is significantly higher than the amplitude of the first overtone pulsation. Therefore, fundamental mode is able to saturate the pulsational instability of both modes. Its limit cycle is firmly stable. In PP models, amplitude of the fundamental mode is strongly reduced and it is no longer able to saturate the pulsation instability alone. This allows the first overtone to grow, and consequently double-mode pulsations arise. When convectively stable regions are treated properly and buoyant forces are included, no stable double-mode behaviour can be found (at least in the parameter range considered in this analysis).

Most of the convective double-mode models published up to date, were computed with the Florida-Budapest hydrocode. In this code PP convection model was adopted: buoyant forces as well as convective flux were neglected in convectively stable zones. This assumption however, was never discussed by the Florida-Budapest group. One case of double-mode pulsation in RR~Lyrae models was published by Feuchtinger (1998). The model was computed with the Vienna pulsation hydrocodes (Wuchterl \& Feuchtinger 1998, Feuchtinger 1999). Although it was not explicitly stated, that turbulent source function was neglected in convectively stable regions in the Vienna hydrocode, we suspect that it was so. Our believe is based on the results concerning first overtone Cepheid models computed with both Florida-Budapest and Vienna hydrocodes (Feuchtinger, Buchler \& Koll\'ath 2000). These authors state, that both codes give essentially the same results. If different treatments of source function in convectively stable zones are adopted in these codes, we would expect significant differences in the computed amplitudes of the single-mode models, as we have shown in Paper~I. No such differences were reported in this paper. This indicates that also the double-mode RR~Lyrae model computed by Feuchtinger (1998) may be physically not correct, although computations of RR~Lyrae models, similar to presented in this analysis are necessary to confirm this suspicion. Many convective models of both RR~Lyrae stars and Cepheids were published by Italian group (\eg Bono \& Stellingwerf 1994, Bono, Marconi \& Stellingwerf 1999). We do not know whether any systematic search for double-mode behaviour was performed with this code. However, to our best knowledge, no nonlinear double-mode models were published. The source function in the Italian code has different form, as the convection model is based on the Stellingwerf (1982) model. In the original Stellingwerf model it was assumed that $S\sim \sqrt{Y}$ instead of $S\sim Y$ as in the Kuhfu\ss{} model. Such form of the source term, particularly the fact, that in convectively stable regions source term cannot damp the convective motions, was criticized by Gehmeyr \& Winkler (1992). This drawback however, was removed in the Italian code, through setting $S\sim\mathrm{sgn}(Y)\sqrt{|Y|}$ (Bono \& Stellingwerf 1992, 1994). Therefore, Italian code is void of artificial eddy-viscous damping responsible for double-mode behaviour in PP models (see Paper~I, and work integrals \eg in Bono, Marconi \& Stellingwerf 1999). In the light of our analysis, the lack of double-mode models in the computations of Italian group is not surprising. 

Our extensive search for F/O1 double-mode Cepheid models with NN convection model (which include negative buoyancy effects) yielded null result. For 24 studied sets of convective and physical parameters, double-mode behaviour was not found. Instead, an either-or domain, where both F and O1 single-mode pulsations are stable, separates the first overtone pulsation domain at the hot side of the instability strip and fundamental mode pulsation domain at the cool side. What is more important, soon after fundamental mode becomes linearly unstable, its pulsation amplitude grows and strongly exceeds the pulsation amplitude of the first overtone. Consequently, fundamental mode limit cycle is firmly stable, across the significant part of the instability strip. It is the main factor preventing the occurrence of the double-mode behaviour. 

Nevertheless, the F/O1 double-mode Cepheids do exist in nature. Therefore, a question arises, what is missing or what physical effect is treated incorrectly in our pulsation hydrocodes. Double-mode Cepheid models computed with PP convection, although unphysical, may provide a hint. We need a mechanism that differentially reduces the modal amplitudes. It is also likely that this mechanism should act in the deep interior of the model, where properties of the fundamental and first overtone modes do differ. In our opinion, two issues should be considered in more detail. First, the treatment of turbulent convection in our models and second, the opacities.

Turbulent convection models adopted in pulsation hydrocodes are very simplified. At the present moment, due to limited computer resources, only simple one equation models are suitable for extensive nonlinear computations. These models contain many free parameters and very simplified description of non-local phenomena, such as overshooting. Of these theories, the Kuhfu\ss{} model seems most consistent and correct. Unfortunately, as we have shown, it is not able to reproduce the observed modal selection.  

As envelope convection zones are clearly present in classical pulsators, turbulent convection is a necessary component of pulsation hydrocodes. However, it is not sure at the the moment, whether convection is crucial in bringing up the double-mode behaviour at all. Non-resonant beat phenomenon is found in radiative, and numerically robust $\beta$~Cephei models (Smolec \& Moskalik 2007). Also radiative non-resonant double-mode RR~Lyrae models (Kov\'acs \& Buchler 1993) and $\delta$~Cephei models (Smolec, unpublished) were computed. Although, these models were obtained through playing with artificial viscosity, they are just indication of the physics missing in our hydrocodes. This turns our attention into opacities.

Recently there is a growing evidence that the opacity computations as well as solar chemical composition are still uncertain. This concerns specially the heavier elements and opacity computations in the hot temperature regimes, corresponding to iron opacity bump ($T\sim 2\cdot 10^5$K). Solar chemical composition was recently revised by Asplund \etal (2005). With new solar mixture however, the standard helioseismic solar model is in serious trouble (see \eg Montalban \etal 2006). This indicates that solar chemical composition or opacity computations need further revisions. The strength of the opacity bump is still under debate and recent asteroseismic models indicate that enhancement of opacities is desired (Pamyatnykh, priv. comm.). Also, our computations of $\beta$~Cephei models (Smolec \& Moskalik 2007) indicate, that modal selection strongly depends on the opacities being used (OP \vs OPAL opacities). Enhancement of iron-group opacities may have an effect on modal selection for Cepheid models. It affects the convective properties of our models, producing additional convection zone in the deep interior. For models considered in this analysis, that is of masses around 4.5\MS and with solar metallicity, iron opacity bump is too weak to produce a convection zone, and hence eddy-viscous damping in the deep interior. With enhanced iron opacity bump, we expect that convective zone will develop in the deep interior, producing eddy-viscous damping. To check these effects  we have computed additional sequence of models, AZ (Table~1) with all parameters of set A, but with increased metallicity to $Z=0.03$. In these models we observe a convection zone connected with the iron bump. It produces an eddy-viscous damping, that acts on fundamental mode, but has negligible effect on the first overtone. However, this effect is weak, as the convection zone is very narrow, and confined to typically $2-4$ zones. Double-mode solution is not found. 

One may draw conclusions about double-mode pulsators (\eg their parameters, such as mass or metallicity, as well as properties of the stellar systems they are in) through linear as well as nonlinear computations. Our results invalidate the nonlinear studies done with PP convection model. However, conclusions and results obtained with linear analysis of models computed with PP convection are correct. Significant turbulent energies and consequently strong eddy-viscous damping in convectively stable zones of full amplitude models are clearly nonlinear effect. Crucial eddy-viscous terms do not affect the computed static structure of the models. If turbulent flux is turned off, turbulent energies are negligible in convectively stable regions of the static model (see Paper~I). As we describe in Paper~I, static structure computed with both PP and NN convection models is very similar. Hence, computed periods and period ratios agree very well. Also linear growth rates are similar, slightly differing for models with turbulent flux turned on. Therefore, the main outcome of the linear analysis, domains of simultaneous instability of fundamental and first overtone modes, as well as periods and period ratios, obtained with PP convection model are correct. Thus, we do not expect significant changes in the analysis of Petersen diagram in the context of beat Cepheid metallicities (Buchler \& Szab\'o 2007, Buchler 2008), if NN convection model is used instead of PP model.

We have analysed the modal selection in classical Cepheid models, considering fundamental and first overtone modes only. Hence, our conclusions strictly apply to F/O1 double-mode Cepheids. We haven't searched for possible O1/O2 double-mode Cepheid models. We just note, that no systematic survey of such models was performed up to date. According to Buchler \& Koll\'ath (2000), some O1/O2 models were found by Florida-Budapest group, among models computed with P\'eclet correction, that accounts for radiative losses, however, no details had been given. Since we haven't performed any analysis of possible overtone double-mode models, we cannot say, whether the same unphysical mechanism is responsible for these double-mode overtone models or not. We note however, that O1 and O2 are more confined to surface layers of the model. In the deep interior both these modes have significantly smaller amplitudes than the fundamental mode. The difference between O1 and O2 modes is less pronounced. Therefore, unphysical eddy-viscous dissipation, causing the differential reduction of modal amplitudes, crucial in F/O1 double-mode PP models, should play a lesser role (if any) in case of O1/O2 double-mode models. The search of O1/O2 double-mode behaviour with our pulsation hydrocodes (NN convection) is planned.

\Acknow{We are grateful to prof. Wojciech Dziembowski for fruitful discussions and commenting the manuscript. Alosza Pamyatnykh is acknowledged for the permission to use the opacity interpolating subroutines and computation of opacity tables with new solar mixture. This work has been supported by the Polish MNiSW Grant No. 1 P03D 011 30.}

\newpage
\begin{figure}[htb]%1
\includegraphics[width=12cm]{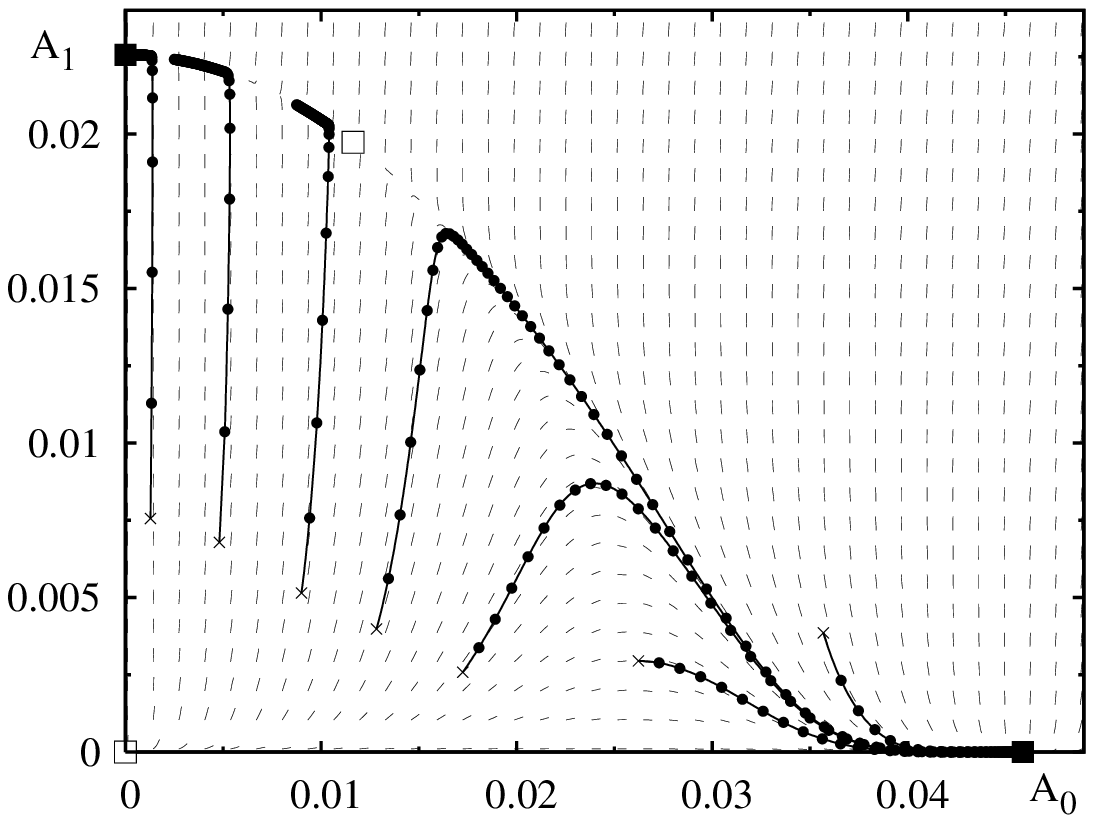}
\FigCap{ Results of hydrodynamic computations for typical Cepheid model. The same static model was initialized with different initial conditions (crosses in the Figure) and time evolution was followed. Solid circles along each trajectory are equally spaced in time, providing information about evolution speed. Solid and open squares mark the location of the computed fixed points, stable and unstable, respectively. Normalized flow field, ($\dot{A_0}, \dot{A_1}$), is represented by thin lines.}
\end{figure}

\begin{figure}[htb]%2
\includegraphics[width=11.5cm]{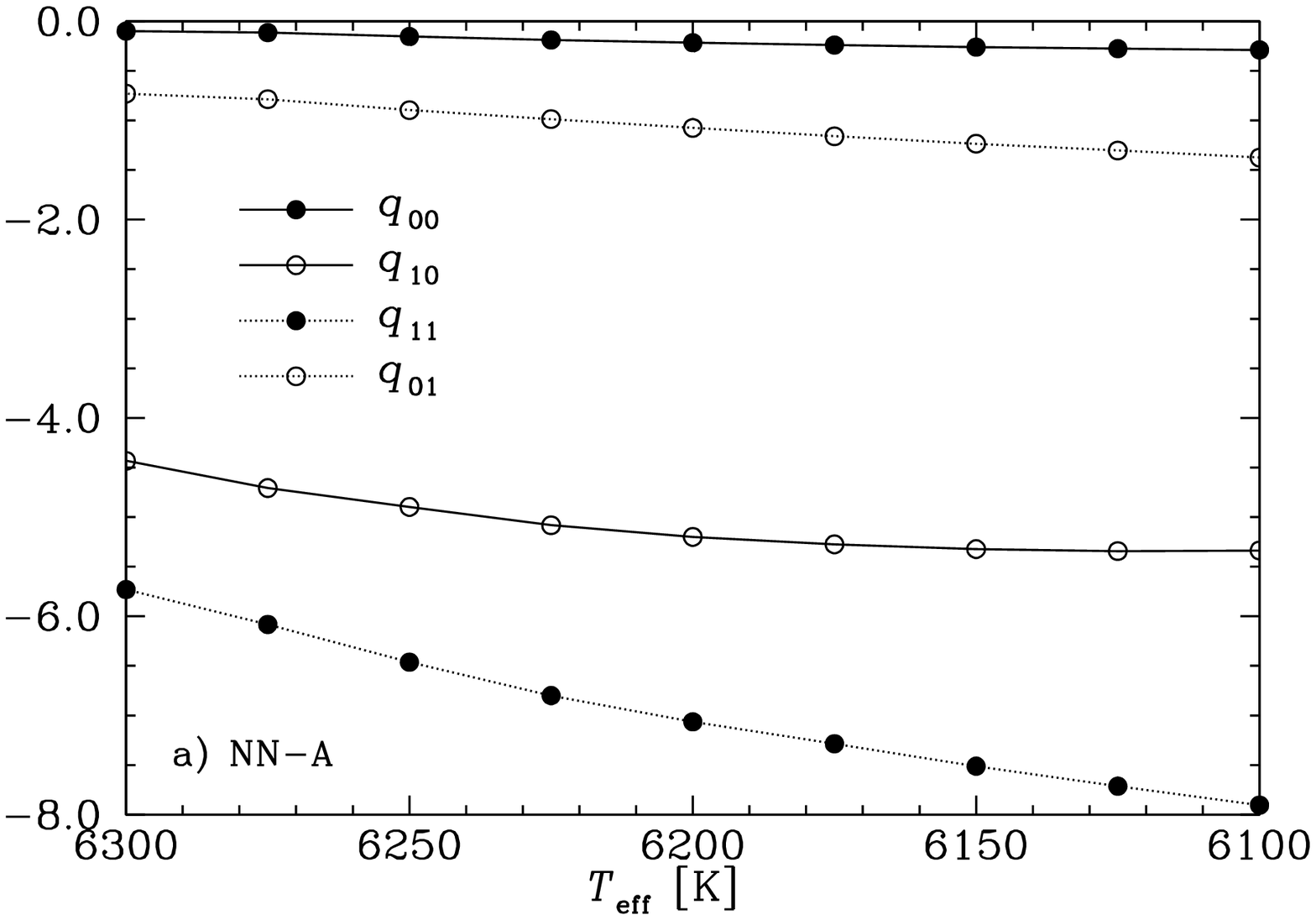}\\
\includegraphics[width=11.5cm]{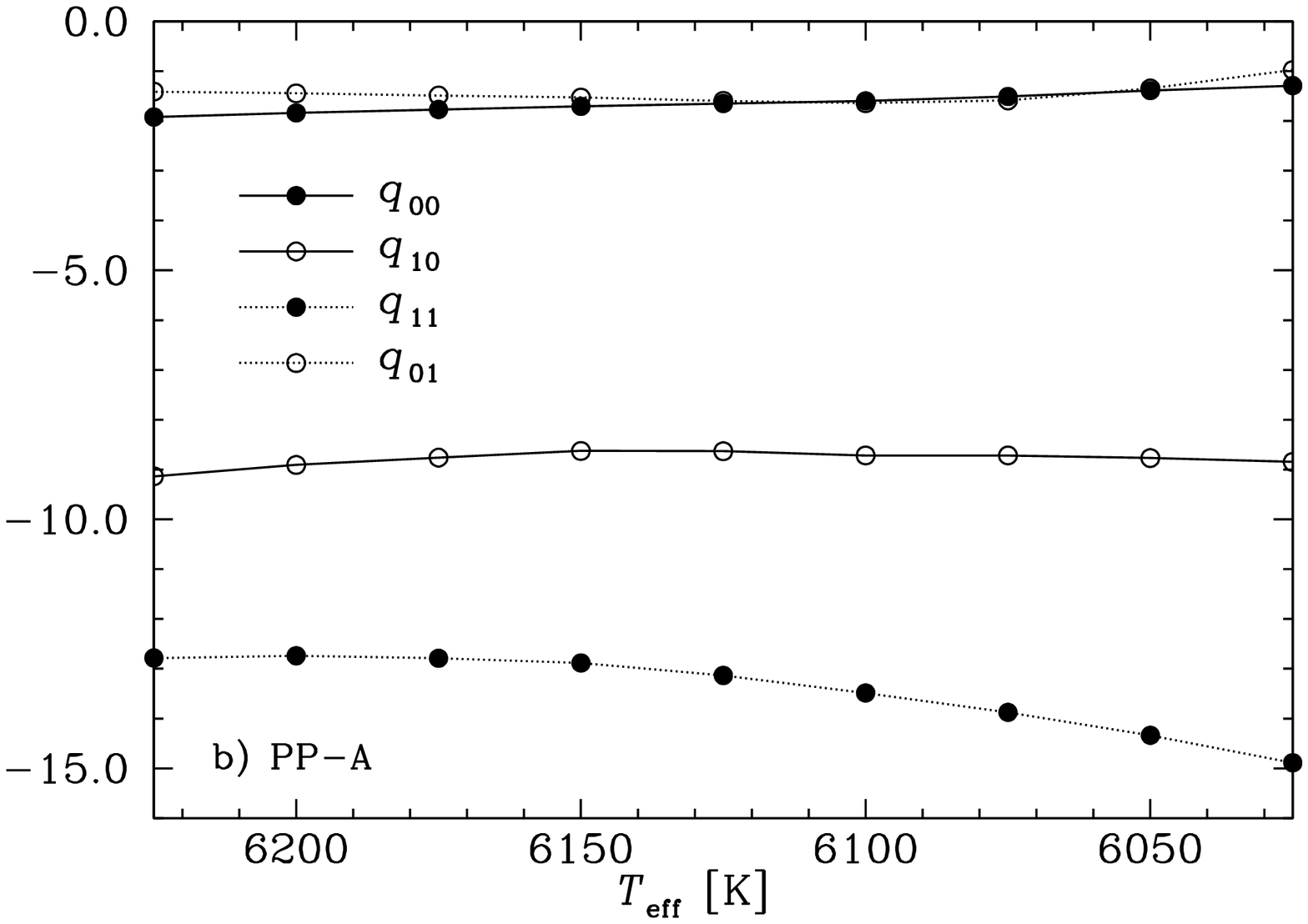}
\FigCap{Cubic saturation coefficients along sequence of Cepheid models of constant luminosity. All hydrodynamic computations done for convective parameters of set A. Either NN convection model (panel a), or PP convection model (panel b) was used. }
\end{figure}

\begin{figure}[htb]%3
\includegraphics[width=11.5cm]{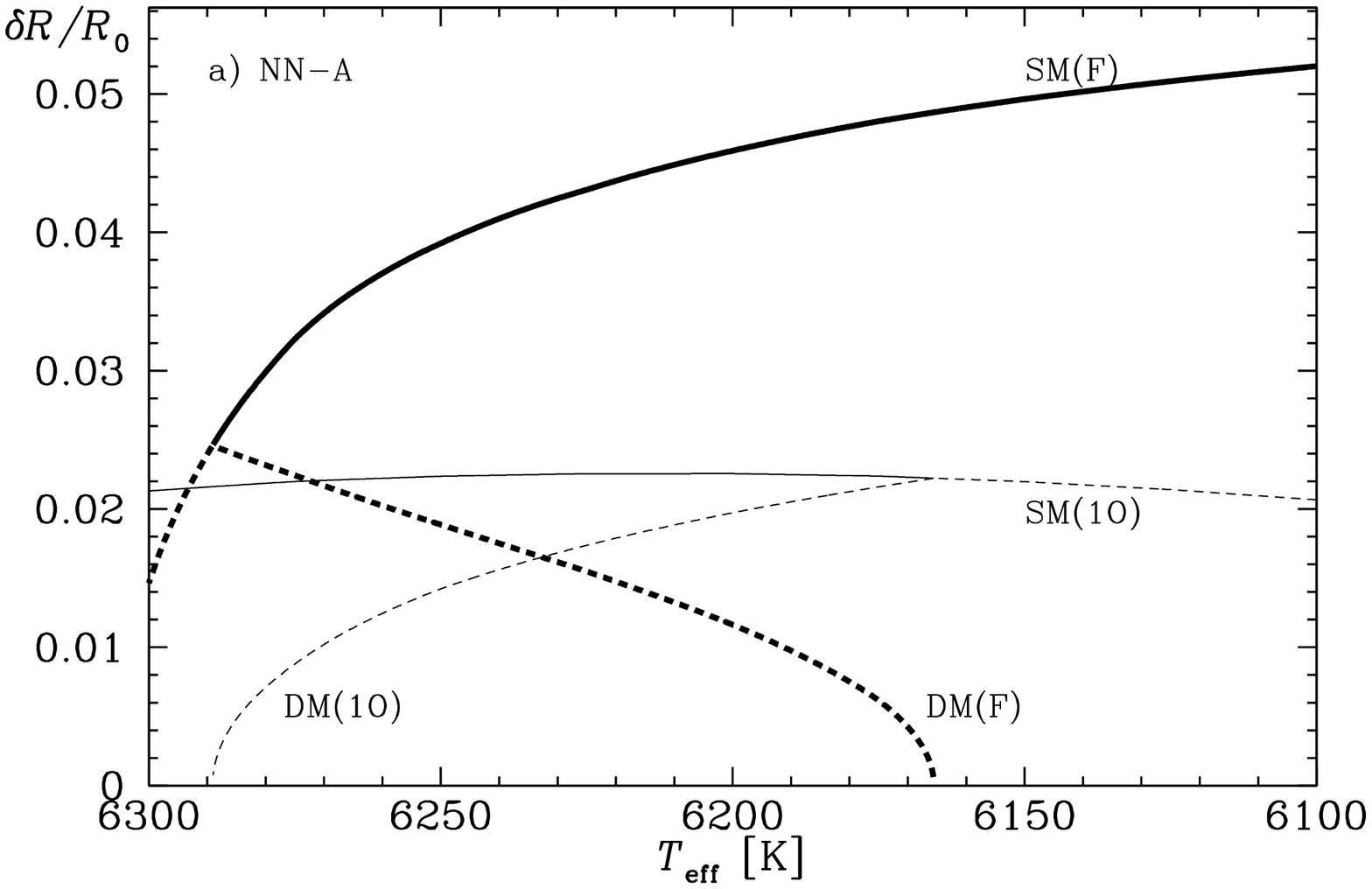}\\
\includegraphics[width=11.5cm]{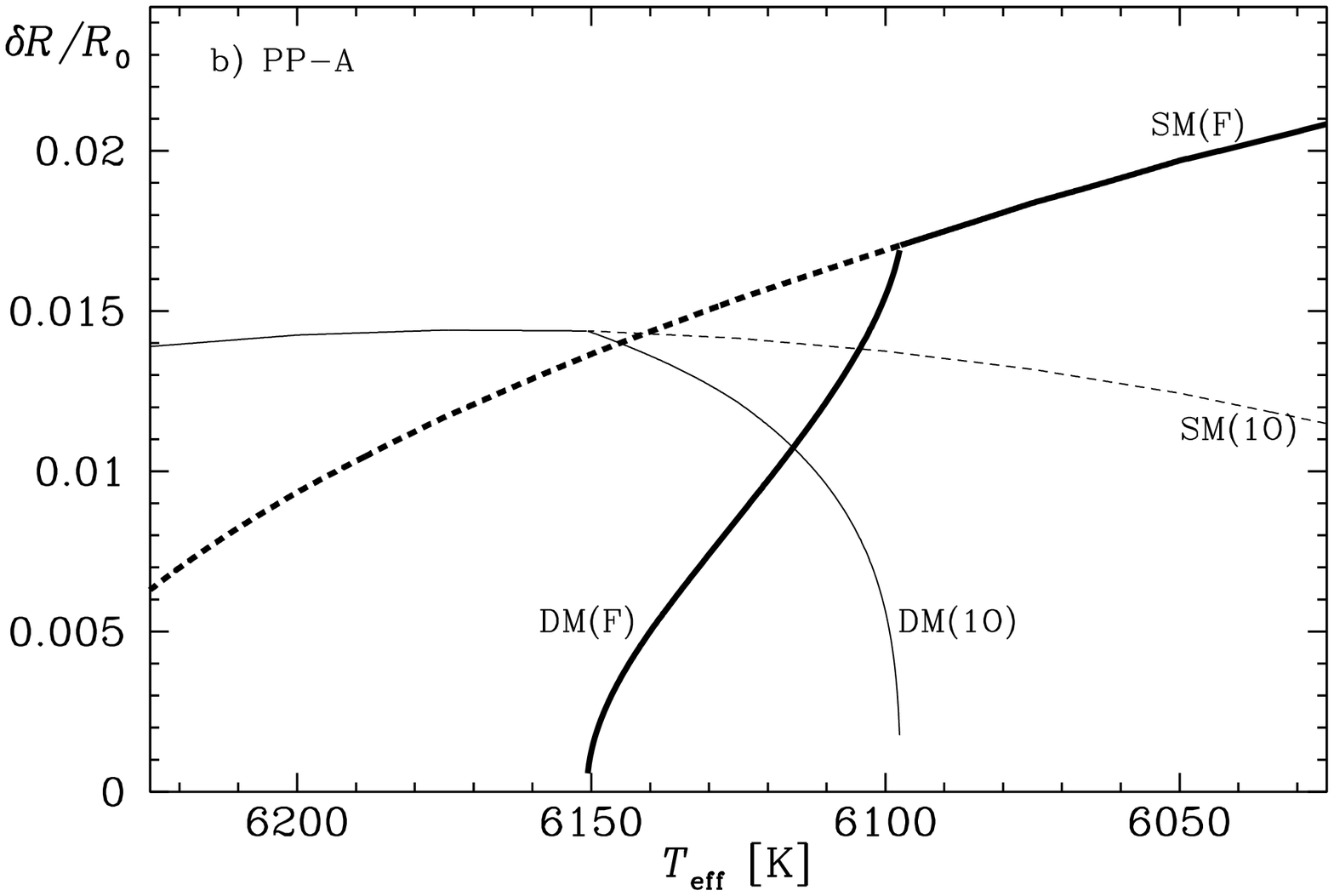}\\
\FigCap{Amplitudes of the single-mode (SM) and double-mode (DM) fixed points along a sequence of Cepheid models of constant luminosity. Solid lines for stable solutions, dashed lines for unstable solutions. Amplitudes of the fundamental mode are plotted with thick lines, while amplitudes of the first overtone mode with thin lines. All hydrodynamic computations done for convective parameters of set A. Either NN convection model (panel a), or PP convection model (panel b) was used.}
\end{figure}

\begin{figure}[htb]%4
\includegraphics[width=12cm]{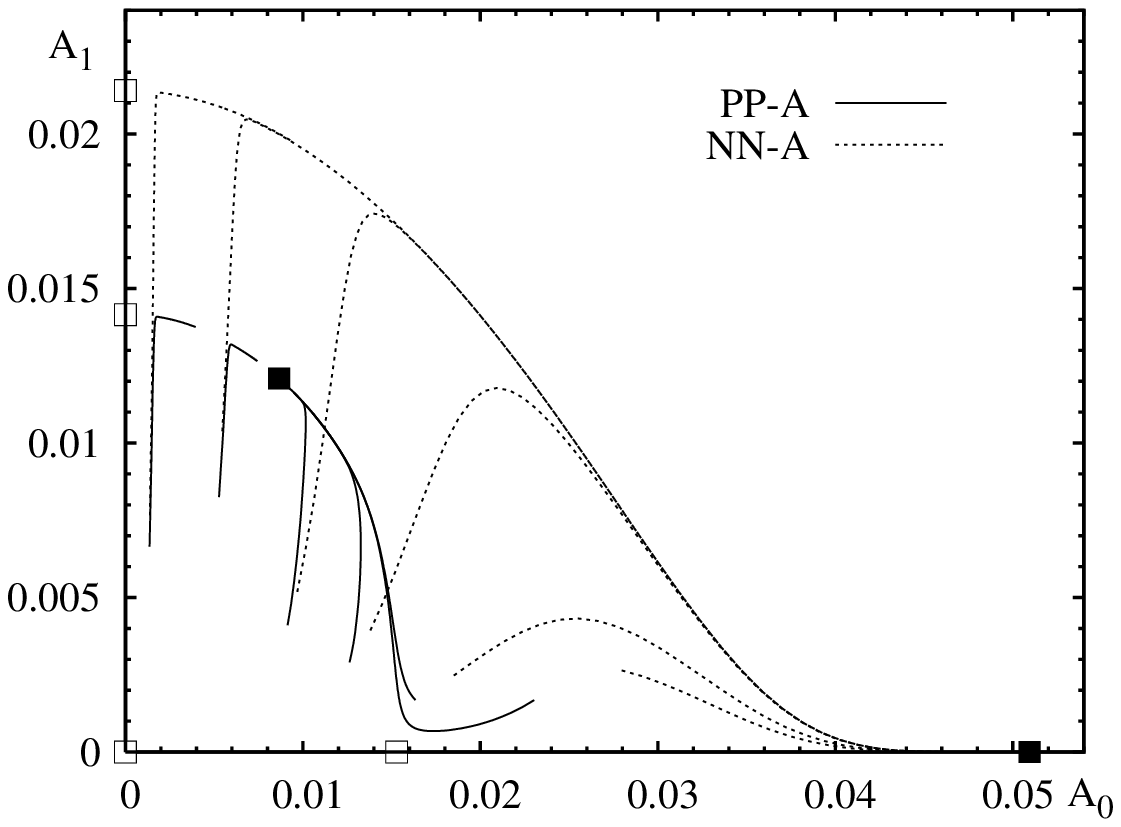}
\FigCap{Results of hydrodynamic computations for Cepheid model of set A ($T_{\mathrm{eff}}=6125$K). Trajectories computed with PP convection model are plotted with solid lines, while trajectories computed wit NN convection model with dotted lines. Solid and open squares mark the location of the computed fixed points, stable and unstable, respectively.}
\end{figure}

\begin{figure}[htb]%5
\includegraphics[width=11.5cm]{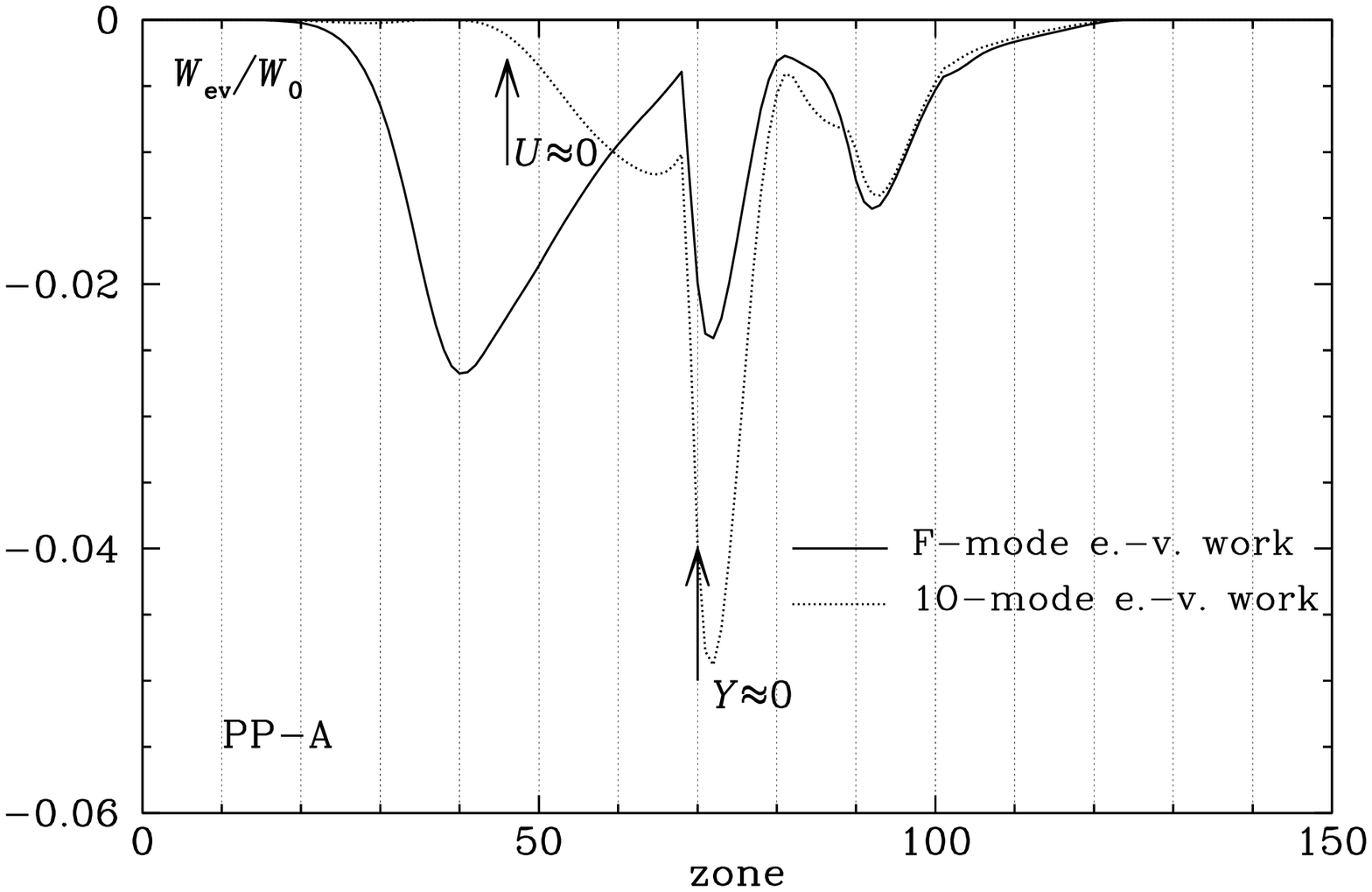}\\
\includegraphics[width=11.5cm]{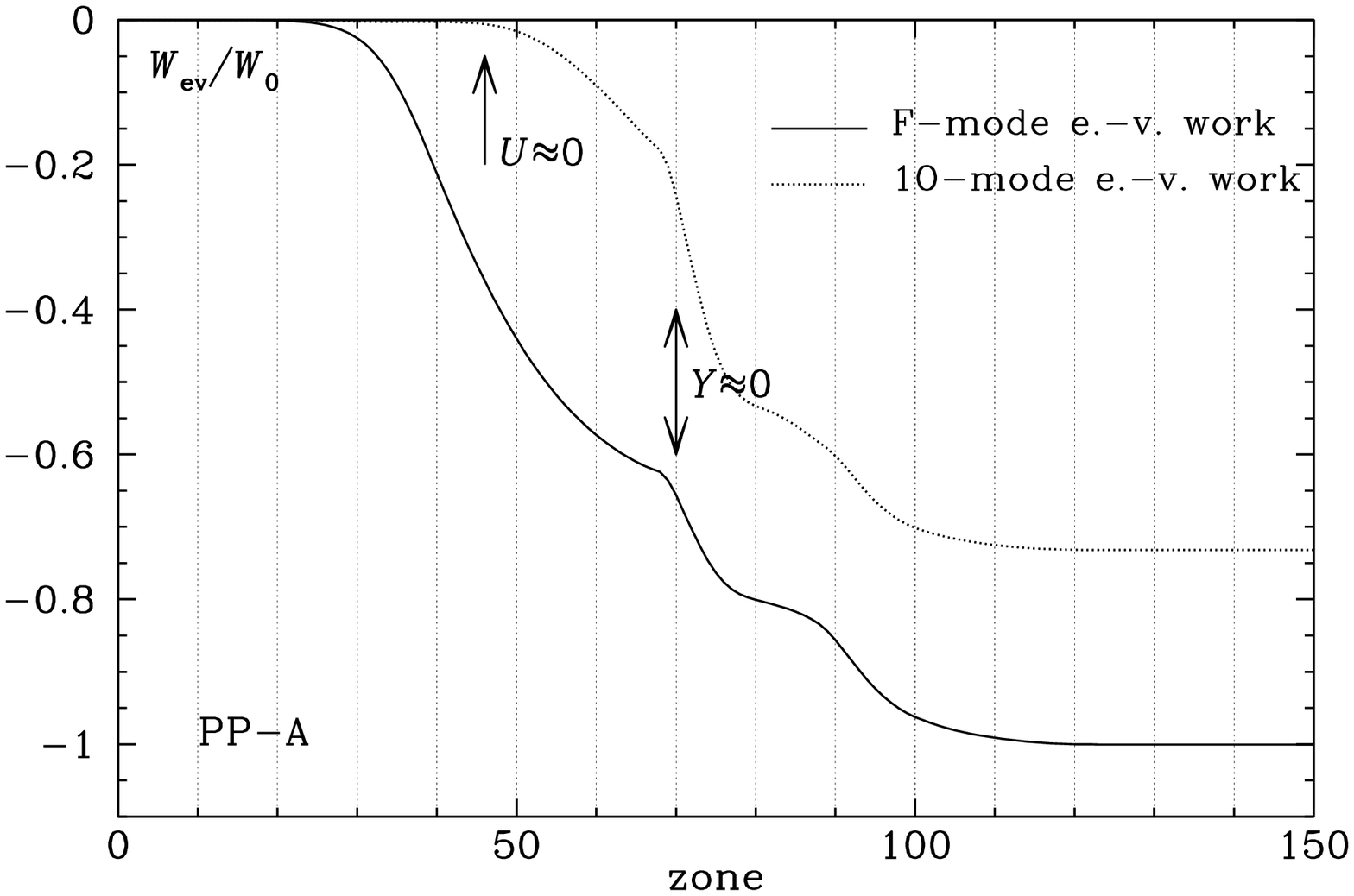}
\FigCap{Normalized, nonlinear eddy-viscous work integrals, local (upper panel) and cumulative (lower panel), for the fundamental mode (solid line) and first overtone (dotted line). PP convection model was used and model parameters of set A. Work integrals are plotted versus the zone number. Surface at right. }
\end{figure}

\begin{figure}[htb]%6
\includegraphics[width=11.5cm]{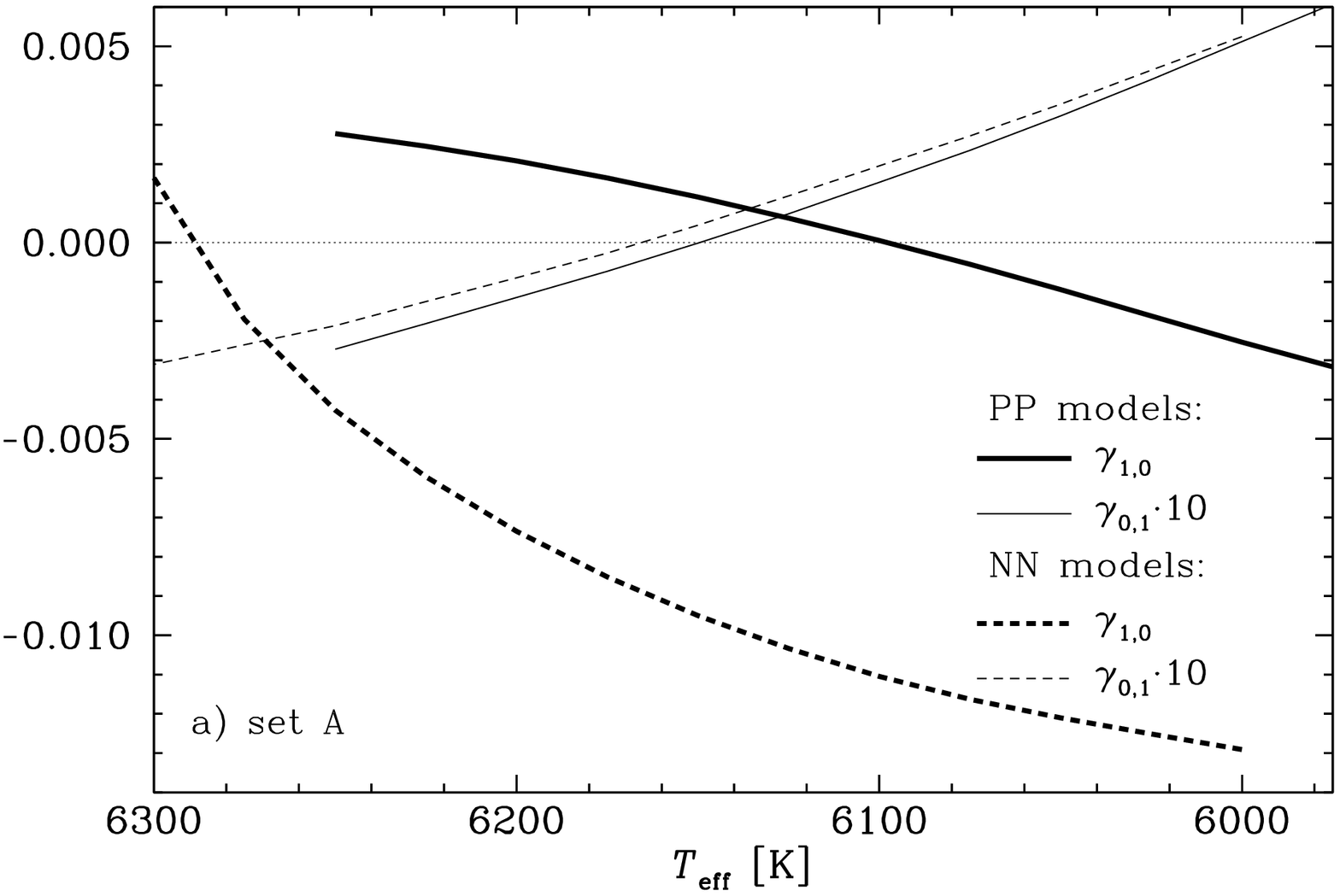}\\
\includegraphics[width=11.5cm]{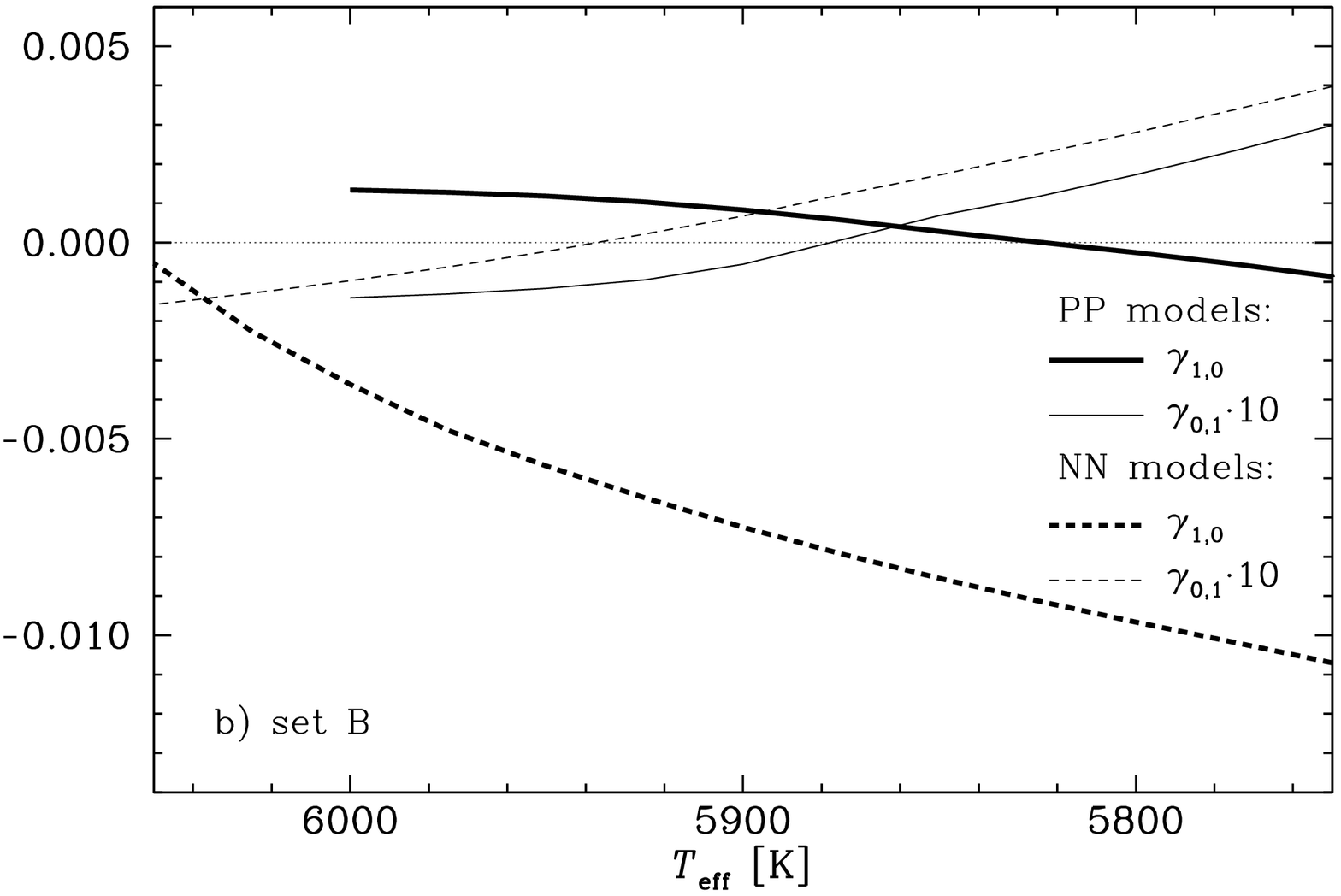}
\FigCap{The run of stability coefficients along a sequence of Cepheid models. Panel a) for models of set A, while panel b) for models of set B.}
\end{figure}

\MakeTable{l|cccccccc|rrrr}{12.5cm}{Values of convective and stellar parameters for the model sequences computed in this paper. $\alpha_p$, $\alpha_s$, $\alpha_c$, $\alpha_d$ and $\alpha_r$ are given in the units of standard values, which are: $\alpha_p=2/3$, $\alpha_s=\alpha_c=1/2\sqrt{2/3}$, $\alpha_d=8/3\sqrt{2/3}$ and $\alpha_r=2\sqrt{3}$ (see Paper I). Masses and luminosities are in the solar units.}
{\hline
Set & $\alpha$ & $\alpha_m$ & $\alpha_s$ & $\alpha_c$ & $\alpha_d$ & $\alpha_p$ & $\alpha_t$ & $\gamma_r$ & $M$ & $L$ & $X$ & $Z$\\ 
\hline
\multicolumn{13}{c}{{\em Basic models:}}\\
\hline
A & 1.5 & 0.20 & 1.0 & 1.0 & 1.0 & 0.0 & 0.00 & 0.0 & 4.5 & 1143.5 & 0.700 & 0.020\\%3
B & 1.5 & 0.25 & 1.0 & 1.0 & 1.0 & 1.0 & 0.01 & 0.0 & \multicolumn{4}{c}{$\ldots$}\\%6t
\hline
\multicolumn{13}{c}{{\em Other models considered:}}\\
\hline
A1 & 1.3 & 0.20 & 1.0 & 1.0 & 1.0 & 0.0 & 0.00 & 0.0 & \multicolumn{4}{c}{$\ldots$}\\%21
A2 & 1.7 & 0.20 & 1.0 & 1.0 & 1.0 & 0.0 & 0.00 & 0.0 & \multicolumn{4}{c}{$\ldots$}\\%22
A3 & 1.5 & 0.25 & 1.0 & 1.0 & 1.0 & 0.0 & 0.00 & 0.0 & \multicolumn{4}{c}{$\ldots$}\\%2
A4 & 1.5 & 0.25 & 0.8 & 1.0 & 1.0 & 0.0 & 0.00 & 0.0 & \multicolumn{4}{c}{$\ldots$}\\%8
A5 & 1.5 & 0.25 & 1.0 & 0.8 & 1.0 & 0.0 & 0.00 & 0.0 & \multicolumn{4}{c}{$\ldots$}\\%9
A6 & 1.5 & 0.25 & 1.0 & 1.0 & 1.2 & 0.0 & 0.00 & 0.0 & \multicolumn{4}{c}{$\ldots$}\\%10
A7 & 1.5 & 0.25 & 0.8 & 0.8 & 1.2 & 0.0 & 0.00 & 0.0 & \multicolumn{4}{c}{$\ldots$}\\%20
A8 & 1.5 & 0.30 & 1.0 & 1.0 & 1.0 & 0.0 & 0.00 & 1.0 & \multicolumn{4}{c}{$\ldots$}\\%11
B1 & 1.5 & 0.20 & 1.0 & 1.0 & 1.0 & 1.0 & 0.10 & 0.0 & \multicolumn{4}{c}{$\ldots$}\\%5t
B2 & 1.5 & 0.20 & 1.0 & 1.0 & 1.0 & 1.0 & 0.50 & 0.0 & \multicolumn{4}{c}{$\ldots$}\\%7t
\hline
AC1& 1.5 & 0.20 & 1.0 & 1.0 & 1.0 & 0.0 & 0.00 & 0.0 & 4.5 &1143.5 & 0.716& 0.010\\%3e
AC2& \multicolumn{8}{c|}{$\ldots$} & 4.5 &1143.5 & 0.756& 0.004\\%3c
AC3& \multicolumn{8}{c|}{$\ldots$} & 4.5 &1143.5 & 0.726& 0.004\\%3d
AC4& \multicolumn{8}{c|}{$\ldots$} & 4.5 &1143.5 & 0.700& 0.012\\%3d
AL1& \multicolumn{8}{c|}{$\ldots$} & 4.5 & 902.5 & 0.700 & 0.020\\%3q
AL2& \multicolumn{8}{c|}{$\ldots$} & 4.5 & 404.4 & 0.700 & 0.020\\%3l
AM & \multicolumn{8}{c|}{$\ldots$} & 4.0 & 751.9 & 0.700 & 0.020\\%3b
BC1& 1.5 & 0.25 & 1.0 & 1.0 & 1.0 & 1.0 & 0.01 & 0.0 & 4.5 &1143.5 & 0.716& 0.010\\%6te
BC2& \multicolumn{8}{c|}{$\ldots$} & 4.5 &1143.5 & 0.756& 0.004\\%6tc
BL1& \multicolumn{8}{c|}{$\ldots$} & 4.5 & 902.5 & 0.700 & 0.020\\%6tq
BL2& \multicolumn{8}{c|}{$\ldots$} & 4.5 & 404.4 & 0.700 & 0.020\\%6tl
BM & \multicolumn{8}{c|}{$\ldots$} & 4.0 & 751.9 & 0.700 & 0.020\\%6tb
\hline
AZ & 1.5 & 0.20 & 1.0 & 1.0 & 1.0 & 0.0 & 0.00 & 0.0 & 4.5 & 1143.5 & 0.700 & 0.030\\%3
\hline
}

\end{document}